\documentclass[11pt]{article}
\usepackage[preprint]{acl}
\usepackage{times}
\usepackage{latexsym}
\usepackage[T1]{fontenc}
\usepackage[utf8]{inputenc}
\usepackage{microtype}
\usepackage{inconsolata}
\usepackage{amsmath,amssymb}
\usepackage{booktabs,array,graphicx}

\title{SecJev: Bringing Security Expertise to System One Decision Models}
\author{
Zheng Chen\textsuperscript{1}, Fei Yu\textsuperscript{2}, Haohao Huang\textsuperscript{1},\\
\bfseries Yang Li\textsuperscript{2}, Anlong Chen\textsuperscript{1}, Lei Chen\textsuperscript{2}\\[0.4em]
\normalfont\textsuperscript{1}University of Electronic Science and Technology of China\\
\normalfont\textsuperscript{2}National Key Laboratory of Security Communication\\[0.4em]
\normalfont\small\url{https://github.com/UESTC1010/SecJev}
}
\hypersetup{
  pdftitle={SecJev: Bringing Security Expertise to System One Decision Models},
  pdfauthor={Zheng Chen, Fei Yu, Haohao Huang, Yang Li, Anlong Chen, Lei Chen}
}

\begin{document}
\maketitle

\begin{abstract}
Security workflows need models that turn complex observations and explicit policies into decisions. System One models introduced by Jev return typed predictions and probabilities; security specialization supplies the domain expertise behind those predictions. We introduce SecJev, to our knowledge the first family of Jev-style decision models specialized for security, spanning 0.8B to 9B parameters. Built on Kev's single-pass candidate scorer, SecJev learns Boolean, choice, and ordered decisions from text, telemetry, and observation histories. We develop SecJev-Corpus to unify source-label prediction and explicit-policy evaluation across 14 tasks and eight sources. It covers tool outputs, traffic, federated updates, consensus, authentication, and vehicle messages. Scene-weighted training adapts the models across these domains while preserving a shared typed decision interface. Security specialization improves every model in the family; SecJev-0.8B outperforms general Kev-9B by 20.51 percentage points in task-macro accuracy. At 0.8B and 4B, SecJev achieves accuracy and latency close to answer-only generative fine-tuning with lower peak allocated GPU memory. SecJev-4B improves on Kev-4B and security-trained generation in prompt-injection and traffic transfer tests. We release adapters, decision heads, SecJev-Corpus, and training and inference code.
\end{abstract}

\section{Introduction}
Security workflows repeatedly turn complex observations into decisions about tool outputs, authentication histories, or voting evidence \citep{debenedetti2024agentdojo,akent-2015-enterprise-data,bano_et_al:LIPIcs.OPODIS.2021.7}. The same tool response can require both injection detection and a trust-boundary action, while voting evidence can require both a quorum judgment and a policy decision. Each answer depends on how the observation relates to the question and any applicable policy.

\begin{figure*}[!t]
\centering
\includegraphics[width=\textwidth]{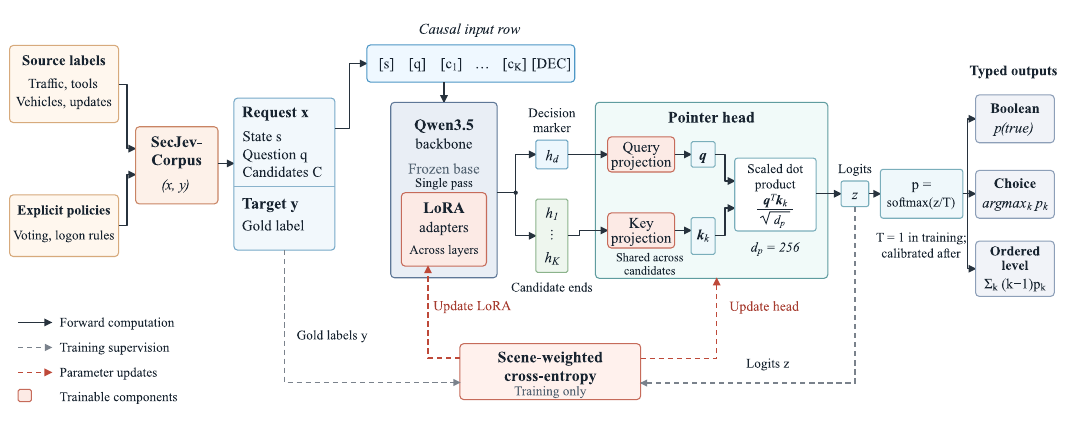}
\caption{SecJev's shared decision path and security adaptation. Source annotations and explicit policies provide requests and targets. One backbone with layerwise LoRA adapters produces decision-marker and candidate-end states; the pointer head scores candidates to obtain typed outputs. Solid edges carry forward computation during both training and inference. Gray dashed edges supply gold labels and logits to scene-weighted cross-entropy; coral dashed edges update only LoRA and the head, leaving the backbone frozen. General-decision initialization uses official Kev for 0.8B/4B/9B and our pretrained 2B model.}
\label{fig:method}
\end{figure*}

Jev introduced System One models in September 2026 that return typed decisions and probabilities directly \citep{almeida2026systemone}. Kev provides open weights for training and running Jev-style models \citep{palmer2026kev}. Here, \emph{Jev-style} denotes models that take observations, a question, and candidate meanings as input and return probabilities over Boolean, choice, or ordered alternatives. Security specialization equips this general decision interface with the expertise to interpret domain observations and apply stated policies.

We introduce \textbf{SecJev}, to our knowledge the first family of Jev-style models specialized for security, at 0.8B, 2B, 4B, and 9B. SecJev brings security expertise to Kev's single-pass candidate scorer through trained adapters and a decision head. All four models support Boolean, choice, and ordered outputs, allowing distinct questions about a scene to share the same model and typed interface.

We develop \textbf{SecJev-Corpus} to train this capability across tool outputs, traffic, federated updates, consensus, authentication, and vehicle messages. Its 170,185 questions combine source-annotation prediction from observations with explicit-policy evaluation against visible evidence. Questions and candidate descriptions preserve the target semantics of each task. Scene-weighted training balances scenes with different numbers of questions while retaining the single-pass decision path (Figure~\ref{fig:method}).

Security specialization improves all 14 tasks at all four sizes. SecJev-0.8B gains 39.40 percentage points in task-macro accuracy over Kev-0.8B and exceeds general Kev-9B by 20.51 points. At 0.8B and 4B, SecJev combines accuracy and latency close to answer-only generation with lower peak allocated GPU memory. General-decision initialization improves the 0.8B model over training the pointer directly on Qwen3.5. Four-shot BIPIA and IoT-23 tests extend the gains to related security tasks. Paraphrase and candidate-permutation tests show stable task aggregates, while general-task evaluation measures capability retention. We release the models, SecJev-Corpus, and code.

\section{Related Work}
\paragraph{Domain and security specialization.}
Domain-adaptive pretraining and instruction tuning focus language models on specialized distributions \citep{gururangan-etal-2020-dont,wadden-etal-2025-sciriff,shi-etal-2025-llama,zeng-etal-2024-agenttuning}. CyberBench combines a multitask cybersecurity benchmark with CyberInstruct, a family of security instruction-tuned generative models \citep{liu2024cyberbench}. SecBench evaluates security knowledge and reasoning through multiple-choice and short-answer questions \citep{jing2024secbench}. LLM fine-tuning also supports anomaly detection in behavior logs \citep{song-etal-2025-confront}. SecJev develops a security-specialized Jev-style family whose shared scorer conditions on the observation, question, and candidate meanings. This supports both source-label prediction and explicit-policy decisions across security domains.

\paragraph{Decision interfaces.}
Jev introduced typed System One predictions \citep{almeida2026systemone}; Kev supplies the pointer head and Boolean, choice, and ordered-score interfaces used here \citep{palmer2026kev}. SecJev specializes this architecture through low-rank adaptation (LoRA) \citep{hu2022lora}, with temperature scaling for confidence \citep{pmlr-v70-guo17a}. Prompt-injection attacks on Jev motivate evaluating how decision models interpret security evidence \citep{wu2026decision}. Classification-head fine-tuning also provides discriminative multiple-choice decisions, scoring each context--candidate sequence separately \citep{sood2026punching}. SecJev brings security specialization to Jev-style models with a shared single-pass candidate scorer, and compares the family with one-token generation and candidate-sequence classification.

\paragraph{Security observations and policies.}
AgentDojo and InjecAgent provide tool-integrated injection scenarios \citep{debenedetti2024agentdojo,zhan-etal-2024-injecagent}. CICIoT2023, ToN-IoT, LANL, and VeReMi supply network, authentication, and vehicle observations \citep{neto2023ciciot,moustafa2021network,kent-2015-cyberdata1,hermann2026veremi}. Twins and ByzFL support controlled consensus and federated-learning scenarios \citep{bano_et_al:LIPIcs.OPODIS.2021.7,gonzalez2025byzfl}. SecJev-Corpus preserves their target semantics while expressing questions and candidate meanings through one interface.

\section{SecJev}
\label{sec:method}
\subsection{Decision Interface}
A scene contains a state $s$ and one or more questions. Each question supplies an instruction $q$ and candidate descriptions $C=(c_1,\ldots,c_K)$. The state holds text, telemetry, or an observation history, including any policy required by the question. A Boolean request returns the probability of \texttt{true}; a choice returns candidate probabilities and the highest-scoring key; an ordered request additionally returns the expected level. Candidate descriptions are inputs, allowing one head to support different answer vocabularies.

Kev renders each question as an independent causal row containing the state, question, delimited candidates, and a final decision marker, preserving field names and order. This accommodates Qwen3.5's hybrid recurrent and attention backbone \citep{qwen3.5}. Batches contain independently encoded questions, with candidate order visible within each row.

\subsection{Single-Pass Candidate Scoring}
Let $h_d$ be the hidden state at the final decision marker and $h_k$ the state at candidate $k$'s closing delimiter. Kev's learned affine projections map both to dimension $d_p=256$, giving
\begin{equation}
 z_k=\frac{(W_Q h_d+b_Q)^{\mathsf T}(W_K h_k+b_K)}{\sqrt{d_p}}.
\end{equation}
The projections are shared across candidates. Probabilities are
\begin{equation}
 p(c_k\mid s,q,C)=\frac{\exp(z_k/T)}{\sum_{j=1}^{K}\exp(z_j/T)},
 \label{eq:probability}
\end{equation}
with $T=1$ during training. Prediction takes the maximum-probability candidate. For levels $0,\ldots,K-1$, the expected level is $\sum_k(k-1)p(c_k\mid s,q,C)$. Answer keys and probability fields are assembled after scoring; no answer tokens are decoded autoregressively.

\subsection{Security Adaptation}
The 0.8B, 4B, and 9B models inherit released Kev adapters and heads. The 2B model first learns general decisions on Qwen3.5-2B-Base using Kev's general-decision data and a supplement covering date reasoning and missing evidence. All four use a three-epoch security-training budget.

We freeze the backbone and train rank-16 LoRA adapters with scaling factor 32 and dropout 0.05, together with the pointer head. Adapters cover attention, feed-forward, and DeltaNet projections. Every security question appears once per epoch, without general-domain replay. To balance scenes containing different numbers of questions, question $j$ in scene $i$ receives weight $1/m_i$. For $N$ scenes, the objective is
\begin{equation}
 \mathcal{L}=-\frac{1}{N}\sum_{i=1}^{N}\frac{1}{m_i}
 \sum_{j=1}^{m_i}\log p(y_{ij}\mid s_i,q_{ij},C_{ij}).
\end{equation}
All output types share this candidate cross-entropy objective.

A separate calibration split fits one positive scalar temperature by minimizing negative log-likelihood (NLL). Scaling changes confidence while preserving the highest-scoring candidate. Appendix~\ref{app:checkpoint-choice} specifies checkpoint selection and Appendix~\ref{app:implementation} gives the training details.

\section{SecJev-Corpus}
\label{sec:corpus}
SecJev-Corpus contains 170,185 questions from 115,324 scenes and eight sources (Table~\ref{tab:tasks}). There are 137,067 Boolean, 25,721 choice, and 7,397 ordered questions. The ordered task concerns consensus evidence, where each level has an explicit meaning. Construction follows two target families: source-annotation prediction from observations, and policy evaluation from a displayed rule and history. Questions and candidate descriptions specify which decision the shared model should make.
\begin{table*}[!t]
\centering
\small
\begin{tabular*}{\textwidth}{@{}@{\extracolsep{\fill}}lllr@{}}
\toprule
Source & Decision task & Type & Questions \\
\midrule
CICIoT2023 & Attack-capture label & Boolean & 48,000 \\
ToN-IoT & Connection attack label & Boolean & 40,000 \\
Twins / Streamlet & Observed candidate-voter conflict & Boolean & 7,397 \\
 & Candidate quorum & Boolean & 7,397 \\
 & Certificate policy action & Choice & 7,397 \\
 & Evidence level & Ordered & 7,397 \\
VeReMi NextGen & Message-manipulation label & Boolean & 11,249 \\
 & Ten-way attack/normal label & Choice & 11,249 \\
ByzFL-based experiments & Attack-generated client update & Boolean & 17,000 \\
LANL authentication & Prior-failure threshold & Boolean & 3,951 \\
 & Logon policy triage & Choice & 3,951 \\
AgentDojo & Embedded prompt injection & Boolean & 2,073 \\
 & Trust-boundary action & Choice & 2,070 \\
InjecAgent & Requested operation & Choice & 1,054 \\
\midrule
Total & 14 tasks & & 170,185 \\
\bottomrule
\end{tabular*}
\caption{Corpus coverage across all splits. Several tasks can share a scene; the 17,000 client-update questions instantiate one task.}
\label{tab:tasks}
\end{table*}

\subsection{Task Construction}
\paragraph{Tool-output security.}
AgentDojo scenes preserve the trusted request, tool call, untrusted response, and trust policy. Questions ask whether an embedded injection is present and whether the response should be used as data or its redirecting instruction excluded. InjecAgent questions distinguish instructions requesting information transfer from other operations. The target describes the requested operation, not attack success or the sensitivity of every transmitted item. A detection formulation with only positive labels is excluded.

\paragraph{Traffic and vehicle observations.}
CICIoT2023 predicts attack-capture labels from measured traffic. ToN-IoT predicts connection labels from protocol and application telemetry, omitting endpoint identities. VeReMi uses local vehicle-message histories to predict the final message's manipulation label and one of ten attack/normal categories. These tasks predict the original source annotations from the displayed observations.

\paragraph{Client updates and voting.}
ByzFL-based supervision comes from 80 controlled MNIST experiments. Inputs contain measured update magnitudes, median distances, and historical cosine similarities, informed by FoolsGold \citep{Fung2020}. Labels distinguish attack-generated updates from honest local training, including non-IID clients. Twins scenes show received messages, a candidate block and round, validators, and a quorum threshold. Questions address visible conflict, quorum, certificate action, and evidence level. The stated policy rejects an observed conflicting voter, otherwise accepts a quorum, and otherwise waits. The targets therefore describe the received evidence and its policy consequences.

\paragraph{Authentication policies.}
LANL scenes use the current event and a source account/computer's preceding 300-second history \citep{akent-2015-enterprise-data}. Same-second events are excluded because their ordering is unknown. Questions concern a prior-failure threshold and a stated logon-triage policy. This is post-event triage: the current success/failure may be visible. This supplies policy-defined supervision for events without attack annotations.

\subsection{Splits and Quality Control}
Records separate visible evidence from labels and provenance. Source labels map to semantic candidates; policy answers are computed from the displayed rule and history. Identifiers are locally aliased without breaking relationships. Questions are added only when the source supports the corresponding distinction.

Table~\ref{tab:splits} gives the four splits. All questions from a scene remain together. Grouping uses captures for CICIoT2023, source hosts for ToN-IoT, simulation runs for Twins/ByzFL, and chronological days for LANL. AgentDojo groups link records sharing normalized tool responses; InjecAgent groups records by trusted user task. VeReMi retains its independent test boundary and groups related non-test vehicle observations. Every semantic class has at least 50 questions per task/split; no class exceeds 80\%, and the largest cross-split class-share gap is 8.58 percentage points. Unordered correct-answer positions are distributed within classes and splits; ordered levels retain their order.
\begin{table}[!t]
\centering
\small
\begin{tabular*}{\columnwidth}{@{}@{\extracolsep{\fill}}lrr@{}}
\toprule
Split & Scenes & Questions \\
\midrule
Training & 81,410 & 118,818 \\
Calibration & 9,333 & 14,821 \\
Development & 9,911 & 14,912 \\
Test & 14,670 & 21,634 \\
\midrule
Total & 115,324 & 170,185 \\
\bottomrule
\end{tabular*}
\caption{Corpus partition. Calibration fits temperature; checkpoint selection is specified in Appendix~\ref{app:checkpoint-choice}.}
\label{tab:splits}
\end{table}

Exhaustive checks cover schemas, candidate-label agreement, duplicates, contradictory labels, and group overlap. Independent policy recomputation checks 37,490 LANL/Twins questions. Inputs exclude supervision and provenance, and fit within 8,192 tokens without truncation. The release preserves source-specific terms, including ToN-IoT's academic-use conditions and requested attribution.\footnote{Companion publications: \citet{booij2022ton,alsaedi2020ton,moustafa2020windows,moustafa2020linux,moustafa2019ton,moustafa2019fog,ashraf2021iotbot}.}

\section{Experiments}
\label{sec:experiments}
We evaluate security capability across sizes, accuracy and inference cost against generation, and transfer to related sources. Analyses examine initialization, semantic consistency, and general-decision retention.

\subsection{Experimental Setup}
\label{sec:setup}
\paragraph{Evaluation.}
We evaluate all 21,634 security-test questions. Task-macro accuracy gives each of the 14 tasks equal weight; micro accuracy weights individual questions. Task-macro balanced accuracy additionally averages class recall within each task. All three output types use their highest-probability candidate for accuracy. A majority baseline selects each task's most frequent training semantic class, independent of answer position. Paired 95\% intervals use 2,000 source-stratified bootstrap draws of whole groups, sharing group weights across associated tasks (Appendix~\ref{app:statistics}).

\paragraph{Training and precision.}
Security adaptation uses AdamW and rank-16 LoRA. Peak learning rates are $2\times10^{-5}$ for the first epoch and $10^{-5}$ for continuation, with effective batches of 128 and 192. Security-test accuracy uses 32-bit floating-point (FP32) backbones for 0.8B/2B/4B and bfloat16 (BF16) for 9B, with FP32 adapters and heads. Each parent--SecJev pair uses matching precision and complete inputs. Appendix~\ref{app:implementation} gives the optimizer and schedule details.

\subsection{Security Specialization and Model Size}
Table~\ref{tab:security-results} measures the capability gained by security specialization, pairing each SecJev model with its general-decision initialization. Adaptation improves every task at every size. Macro gains are 39.40, 40.83, 26.63, and 24.26 percentage points for 0.8B, 2B, 4B, and 9B. All four macro-gain intervals exclude zero; the 4B gain has a 95\% interval of $[18.59,30.75]$ points. The training-majority baseline reaches 62.89\% macro accuracy. SecJev-0.8B exceeds Kev-9B by 20.51 points, using roughly one eleventh of its nominal parameters.
\begin{table*}[!t]
\centering
\small
\setlength{\tabcolsep}{3pt}
\begin{tabular*}{\textwidth}{@{}@{\extracolsep{\fill}}lr*{9}{r}@{}}
\toprule
& & & \multicolumn{2}{c}{0.8B} & \multicolumn{2}{c}{2B} & \multicolumn{2}{c}{4B} & \multicolumn{2}{c}{9B} \\
\cmidrule(lr){4-5}\cmidrule(lr){6-7}\cmidrule(lr){8-9}\cmidrule(lr){10-11}
Task & $n$ & Maj. & Init. & SecJev & Init. & SecJev & Init. & SecJev & Init. & SecJev \\
\midrule
AgentDojo: boundary & 426 & 77.46 & 51.88 & 80.28 & 40.85 & 92.96 & 54.23 & 99.53 & 61.97 & 98.83 \\
AgentDojo: injection & 426 & 77.46 & 32.86 & 79.81 & 29.81 & 92.72 & 73.00 & 98.36 & 64.55 & 98.83 \\
ByzFL: client update & 800 & 80.00 & 79.62 & 92.62 & 80.00 & 97.50 & 79.88 & 95.88 & 80.25 & 93.88 \\
CICIoT2023: attack & 6,000 & 75.00 & 25.00 & 91.80 & 25.00 & 92.13 & 26.28 & 91.72 & 37.93 & 90.27 \\
InjecAgent: operation & 124 & 56.45 & 44.35 & 100.00 & 51.61 & 98.39 & 62.90 & 100.00 & 57.26 & 100.00 \\
LANL: logon action & 540 & 50.93 & 72.41 & 98.89 & 72.78 & 99.81 & 79.81 & 100.00 & 96.11 & 99.63 \\
LANL: failure threshold & 540 & 75.56 & 80.74 & 99.81 & 99.44 & 100.00 & 99.44 & 100.00 & 99.81 & 100.00 \\
ToN-IoT: attack & 3,500 & 75.00 & 25.00 & 98.03 & 25.00 & 98.97 & 50.66 & 98.80 & 24.83 & 98.89 \\
Twins: evidence level & 659 & 39.15 & 41.12 & 98.79 & 40.06 & 99.85 & 81.18 & 99.54 & 81.94 & 99.85 \\
Twins: policy action & 659 & 39.15 & 47.65 & 98.48 & 39.45 & 100.00 & 71.78 & 99.85 & 88.16 & 100.00 \\
Twins: quorum & 659 & 64.49 & 53.41 & 99.70 & 64.95 & 99.85 & 78.15 & 100.00 & 91.81 & 99.85 \\
Twins: conflict & 659 & 69.80 & 64.04 & 99.24 & 70.11 & 99.70 & 97.12 & 99.85 & 96.97 & 100.00 \\
VeReMi: manipulation & 3,321 & 52.24 & 47.76 & 63.11 & 48.12 & 66.21 & 47.76 & 70.34 & 47.76 & 69.26 \\
VeReMi: attack type & 3,321 & 47.76 & 47.24 & 64.14 & 42.97 & 63.69 & 48.09 & 69.23 & 48.15 & 67.90 \\
\midrule
Micro accuracy & 21,634 & 64.25 & 40.53 & 85.03 & 40.44 & 86.48 & 50.49 & 88.01 & 50.87 & 87.18 \\
Task-macro accuracy & 21,634 & 62.89 & 50.94 & 90.34 & 52.15 & 92.98 & 67.88 & 94.51 & 69.82 & 94.08 \\
Task-macro balanced & 21,634 & 43.57 & 51.86 & 88.00 & 51.34 & 90.31 & 67.16 & 92.42 & 70.36 & 91.27 \\
\bottomrule
\end{tabular*}
\caption{Full security-test accuracy (\%). Maj. predicts the training-majority semantic class. Init. denotes official Kev for 0.8B/4B/9B and our general-decision model for 2B. Checkpoint choices are given in Appendix~\ref{app:checkpoint-choice}. Macro weights tasks equally; balanced macro also balances semantic classes.}
\label{tab:security-results}
\end{table*}

Security specialization produces gains throughout the family, with 4B leading task-macro accuracy (94.51\%) and balanced macro (92.42\%). The capability profile spans three settings: at 4B, task-macro accuracy rises from 63.38 to 99.30\% on the three tool-text tasks, from 50.53 to 85.19\% on the five source-label tasks, and from 84.58 to 99.87\% on the six explicit-policy tasks. Near-perfect policy scores show that the models learn to apply stated rules to visible evidence; strong gains also extend to predicting source annotations and interpreting tool responses. ByzFL peaks at 2B, while both VeReMi tasks remain below 72\%. Vehicle-message classification accounts for 6,642 test questions but only two macro-average terms, explaining the lower micro accuracy. Together, the gains on source-label, tool-text, and explicit-policy tasks show security specialization across distinct target semantics through one candidate-scoring interface.

\subsection{Comparison with Other Decision Interfaces}
\label{sec:generative-comparison}
\textbf{SecGen} is the generative supervised fine-tuning (SFT) baseline on SecJev-Corpus at 0.8B and 4B, using matching Qwen3.5 bases, rank-16 LoRA, scene weighting, and three epochs. It predicts one answer-key token with full-vocabulary cross-entropy and a frozen vocabulary head. At 0.8B, we also train a shared scalar classification head on separately encoded context--candidate sequences \citep{sood2026punching}, using the same security recipe. A direct-Qwen pointer measures candidate scoring without Kev initialization. Appendix~\ref{app:generative} gives protocols and development-based selection.

Timing uses one RTX 3090, a common BF16 backbone with FP32 adapters and heads, and matching attention implementations. Accuracy is recomputed on the full security test with the serial timing implementation for Table~\ref{tab:efficiency}. Batch-one latency includes request parsing and validation, tokenization, device transfer, inference, and answer processing after warm-up, excluding model loading. Each request contains one question, with no prompt-cache reuse. Short inputs have at most 512 tokens and long inputs have 2,048--8,192; each stratum contains 100 fixed development questions. Greedy generative decoding uses an answer-key grammar and stops at key completion. Invalid outputs count as errors. We report medians, pooled P95, serial batch-one throughput, and peak allocated GPU memory over five timing passes.
\begin{table*}[!t]
\centering
\small
\setlength{\tabcolsep}{4pt}
\begin{tabular*}{\textwidth}{@{}@{\extracolsep{\fill}}llrrrrrr@{}}
\toprule
Size & Method & Macro acc. & Short ms & Long ms & P95 ms & Q/s & GiB \\
\midrule
0.8B & SecJev & 90.46 & 79.2 & 164.1 & 256.0 & 7.32 & 2.56 \\
0.8B & SecGen & 90.01 & 80.4 & 167.4 & 260.1 & 7.22 & 3.52 \\
0.8B & Direct-Qwen pointer & 88.12 & 80.4 & 164.6 & 256.7 & 7.27 & 2.56 \\
4B & SecJev & 94.48 & 109.1 & 578.8 & 890.5 & 2.54 & 10.18 \\
4B & SecGen & 94.34 & 116.9 & 599.9 & 910.6 & 2.47 & 12.58 \\
9B & Kev & 69.79 & 169.1 & 861.0 & 1278.5 & 1.73 & 17.11 \\
\bottomrule
\end{tabular*}
\caption{Matched accuracy and inference cost. Accuracy (\%) is re-evaluated at common precision. Short/long latencies are medians; P95 is the pooled 95th-percentile latency. Q/s is mean serial throughput at batch size one, and GiB is peak allocated GPU memory. Each pass measures 100 short and 100 long requests.}
\label{tab:efficiency}
\end{table*}

At FP32 precision, SecGen reaches 90.02/94.35\% task-macro accuracy at 0.8B/4B, versus SecJev's 90.34/94.51\%; balanced macro is 87.79/92.30\% versus 88.00/92.42\%. The 0.8B classifier reaches 89.85\% macro and 86.84\% balanced macro, between SecJev and the direct-Qwen pointer (88.17/83.75\%). Security supervision thus supports several output interfaces, with SecJev leading these task aggregates. At common BF16 precision, both SecJev and the one-token generative baseline require one forward pass, with close latency and serial throughput (Table~\ref{tab:efficiency}). SecJev reduces peak allocated GPU memory from 3.52 to 2.56 GiB at 0.8B and from 12.58 to 10.18 GiB at 4B, reductions of about 27\% and 19\%. The candidate scorer thus supports typed security decisions with a smaller measured memory footprint at comparable inference cost. Section~\ref{sec:initialization} isolates the contribution of Kev initialization at 0.8B; Appendix~\ref{app:timing} reports timing variability.

\begin{table*}[!tp]
\centering\small
\begin{tabular*}{\textwidth}{@{}@{\extracolsep{\fill}}llrrrrr@{}}
\toprule
Dataset & Model & $n$ & Accuracy & Precision & Recall & FPR \\
\midrule
BIPIA & Kev-4B & 1000 & 71.30 & 64.14 & 96.60 & 54.00 \\
BIPIA & SecJev-4B & 1000 & 83.60 & 75.38 & 99.80 & 32.60 \\
BIPIA & SecGen-4B & 1000 & 79.60 & 71.39 & 98.80 & 39.60 \\
\midrule
IoT-23 & Kev-4B & 1000 & 73.00 & 65.54 & 97.00 & 51.00 \\
IoT-23 & SecJev-4B & 1000 & 90.20 & 86.95 & 94.60 & 14.20 \\
IoT-23 & SecGen-4B & 1000 & 84.30 & 76.34 & 99.40 & 30.80 \\
\bottomrule
\end{tabular*}
\caption{Transfer performance of 4B models on BIPIA and IoT-23 (\%), with 1,000 test examples per dataset. SecGen-4B is the answer-only generative baseline. Injected content and malicious traffic are the positive classes for precision, recall, and false-positive rate (FPR).}
\label{tab:fresh-main}
\end{table*}

\begin{table*}[!t]
\centering
\small
\begin{tabular*}{\textwidth}{@{}@{\extracolsep{\fill}}llrrrrr@{}}
\toprule
Change & SecJev size & Pairs & Original acc. & Changed acc. & Consistency & Both correct \\
\midrule
Paraphrase & 0.8B & 1400 & 88.64 & 87.79 & 95.36 & 85.93 \\
 & 2B & 1400 & 90.79 & 89.86 & 96.57 & 88.79 \\
 & 4B & 1400 & 91.64 & 91.14 & 98.36 & 90.57 \\
 & 9B & 1400 & 90.71 & 90.64 & 96.43 & 88.93 \\
\midrule
Choice permutation & 0.8B & 500 & 83.00 & 82.60 & 98.80 & 82.20 \\
 & 2B & 500 & 84.80 & 84.00 & 97.40 & 83.40 \\
 & 4B & 500 & 89.00 & 89.00 & 98.40 & 88.20 \\
 & 9B & 500 & 87.40 & 87.80 & 98.60 & 87.00 \\
\bottomrule
\end{tabular*}
\caption{Paired semantic consistency (\%). Original/changed accuracy uses the same eligible questions: 1,400 for paraphrasing and 500 for permutation. Consistency is semantic answer agreement; both-correct requires both predictions to match the label. Tasks have equal weight within each condition.}
\label{tab:robustness}
\end{table*}

\subsection{Transfer to Related Security Data}
\label{sec:external}
We evaluate transfer to BIPIA-derived prompt-injection detection\footnote{\url{https://github.com/microsoft/BIPIA}} and IoT-23 malicious-flow identification,\footnote{\url{https://www.stratosphereips.org/datasets-iot23}} extending the tool-output and traffic settings represented in SecJev-Corpus. Each dataset contains 1,000 binary test examples, with 500 examples per class. All three 4B models use 4-shot prompts with four labeled demonstrations retrieved from a pool disjoint from the test sets. The demonstrations specify source-specific label meanings and decision criteria, providing a compact way to configure a new security task. Model parameters remain fixed.

On BIPIA, SecJev-4B reaches 83.60\% accuracy, exceeding Kev-4B's 71.30\% and SecGen-4B's 79.60\% by 12.30 and 4.00 percentage points (Table~\ref{tab:fresh-main}). SecJev combines 99.80\% recall with 75.38\% precision. Its false-positive rate falls to 32.60\%, from 54.00\% for Kev and 39.60\% for SecGen. In this test, SecJev combines higher injection recall with fewer false alarms.

On IoT-23, SecJev-4B achieves 90.20\% accuracy, versus 73.00\% for Kev-4B and 84.30\% for SecGen-4B, gaining 17.20 and 5.90 points. Precision rises to 86.95\%, and the false-positive rate falls to 14.20\%, versus SecGen's 76.34\% and 30.80\%. SecGen has higher recall (99.40\% versus 94.60\%), while SecJev's fewer benign-flow errors yield higher overall accuracy. SecJev leads both baselines across the two tasks under this four-example setup. These gains support using one security-specialized decision interface for both untrusted text and structured traffic observations.

\section{Analysis}
\label{sec:analysis}
\subsection{Question Meaning and Candidate Order}
A shared decision model should track question and candidate meanings across equivalent formulations. We test paraphrased questions and permuted unordered alternatives, keeping evidence, field order, rules, and semantic labels fixed. We use 100 test questions per task, each paraphrased once (1,400 pairs). Each of 100 questions from five choice tasks also receives one nonidentity permutation (500 pairs); ordered levels are not shuffled. Together with the originals, this yields 3,300 requests per model. Predictions are mapped back to semantic keys.

In Table~\ref{tab:robustness}, paraphrasing reduces accuracy by 0.07--0.93 percentage points across all four sizes, while permutation changes it by $-0.80$ to $+0.40$ points. Answer agreement is 95.36--98.36\% for paraphrases and 97.40--98.80\% for permutations. Both-correct scores separate preserved correct decisions from stable mistakes. Aggregate accuracy is stable across the tested formulations; paired agreement measures which individual decisions persist.

Task-level results separate aggregate stability from changes in individual decisions. For 0.8B, paraphrasing reduces federated-update accuracy from 91\% to 85\% and vehicle attack-type accuracy from 39\% to 32\%. For 9B, ToN-IoT falls from 95\% to 89\%, while CICIoT rises from 88\% to 95\%; these offsetting shifts yield a net loss of just one answer. Across sizes, 15--38 previously correct answers become incorrect among the 1,400 paraphrase pairs. For 9B VeReMi manipulation, prediction agreement is 77\%, with a one-point accuracy change. Fixed-batch controls are discussed in Appendix~\ref{app:robust-details}.

\subsection{Value of General Decision Initialization}
\label{sec:initialization}
The 0.8B comparison fixes the backbone, pointer architecture, security data, and three-epoch budget. One model inherits Kev's adapters and head; the other learns these components directly on Qwen3.5. Development selection favors epoch 3 for both. Kev initialization reaches 90.34\% test macro accuracy versus 88.17\% for direct-Qwen training, a 2.17-point gain; balanced macro rises from 83.75\% to 88.00\% (Table~\ref{tab:initialization}). Direct-Qwen training performs better on AgentDojo boundary actions and federated-update classification, while Kev initialization yields its largest gains on vehicle manipulation and attack type (Appendix~\ref{app:qwen-init}). Overall, general-decision initialization provides a stronger starting point for security adaptation under the fixed training budget. 
\begin{table}[!t]
\centering
\small
\begin{tabular*}{\columnwidth}{@{}@{\extracolsep{\fill}}lrrr@{}}
\toprule
0.8B initialization & Dev. bal. & Test macro & Test bal. \\
\midrule
Kev & 90.36 & 90.34 & 88.00 \\
Qwen3.5 directly & 84.85 & 88.17 & 83.75 \\
\bottomrule
\end{tabular*}
\caption{Initialization comparison (\%). Dev./test bal. are task-macro balanced accuracy. Both initializations select epoch 3 on development data. Test results cover all 21,634 questions.}
\label{tab:initialization}
\end{table}

\subsection{Retention and Confidence}
\label{sec:retention}
We evaluate general capability retention on two Kev suites (Table~\ref{tab:retention}). General decision tests held-out examples from Kev's training sources; general transfer tests new sources and policy-rule types. General decision accuracy changes by at most 0.50 points after security training. General transfer falls from 68.45\% to 66.01\% at 0.8B, while changes at larger sizes stay within 0.46 points. The 0.8B transfer decline has a paired 95\% interval of $[-4.73,-0.15]$ points; all other general-accuracy intervals include zero (Appendix~\ref{app:statistics}).
\begin{table}[!t]
\centering
\small
\setlength{\tabcolsep}{4pt}
\begin{tabular*}{\columnwidth}{@{}@{\extracolsep{\fill}}lrrrr@{}}
\toprule
& \multicolumn{2}{c}{General decision} & \multicolumn{2}{c}{General transfer} \\
\cmidrule(lr){2-3}\cmidrule(lr){4-5}
Size & Init. & SecJev & Init. & SecJev \\
\midrule
0.8B & 83.50 & 83.08 & 68.45 & 66.01 \\
2B & 84.00 & 84.50 & 75.91 & 75.91 \\
4B & 87.08 & 86.58 & 83.69 & 83.84 \\
9B & 87.50 & 87.42 & 85.06 & 84.60 \\
\bottomrule
\end{tabular*}
\caption{General-task accuracy (\%) before and after specialization: 1,200 general decision and 656 general transfer questions. Init. denotes the model before security training. Suite definitions appear in Section~\ref{sec:retention}.}
\label{tab:retention}
\end{table}

We evaluate confidence quality alongside accuracy. Temperature scaling improves security-test NLL for all sizes (Table~\ref{tab:calibration}), while Brier score and expected calibration error (ECE) worsen. For example, 9B NLL falls from 0.364 to 0.355 while ECE rises from 1.53\% to 2.48\%. On both general suites, the frozen security temperature improves SecJev NLL, Brier, and ECE at every size, while general-suite NLL remains higher than for the initialization (Appendix~\ref{app:statistics}). Accuracy retention alone therefore misses changes in confidence.
\begin{table}[!t]
\centering
\small
\begin{tabular*}{\columnwidth}{@{}@{\extracolsep{\fill}}llrrr@{}}
\toprule
Size & Probabilities & NLL & Brier & ECE (\%) \\
\midrule
0.8B & Raw & 0.435 & 0.211 & 2.18 \\
0.8B & Scaled & 0.412 & 0.213 & 3.34 \\
2B & Raw & 0.399 & 0.191 & 1.80 \\
2B & Scaled & 0.385 & 0.194 & 3.28 \\
4B & Raw & 0.358 & 0.173 & 1.84 \\
4B & Scaled & 0.344 & 0.177 & 3.56 \\
9B & Raw & 0.364 & 0.181 & 1.53 \\
9B & Scaled & 0.355 & 0.182 & 2.48 \\
\bottomrule
\end{tabular*}
\caption{Security-test confidence metrics, averaged over questions. Lower is better for all three. ECE uses 15 equal-width bins. Scaling uses only the held-out calibration split.}
\label{tab:calibration}
\end{table}

\subsection{Data Validity and Supervision Semantics}
Exhaustive checks find no schema or candidate-label violations in 170,185 questions and no disagreement in 37,490 independently recomputed policy answers. Two human reviewers independently assess 280 questions, 20 per task, and agree on 261 answers (93.21\%). Both match the corpus labels on all 180 tool-text and explicit-policy questions. On the 100 source-label questions, they agree on 81 answers and both rate the visible evidence as sufficient for 28. The review supports the interpretability of tool and policy supervision, while locating the main uncertainty in inferring source annotations from partial observations (Appendix~\ref{app:audit}).

\section{Conclusion}
SecJev brings security expertise to System One decision models through four Jev-style models with a shared typed interface. Training across 14 tasks improves every model and enables SecJev-0.8B to outperform general Kev-9B. SecJev-Corpus combines source-label prediction and explicit-policy evaluation. The 0.8B and 4B models achieve accuracy close to answer-only generation with lower peak allocated GPU memory. SecJev-4B also improves transfer accuracy on BIPIA and IoT-23. General decision initialization also strengthens the 0.8B model beyond security-only training. Question-conditioned scoring lets one model express distinct judgments about the same evidence through a shared security-trained head. SecJev-Corpus provides the security supervision connecting these judgments to source annotations and explicit policy rules. We release SecJev, SecJev-Corpus, and code for security-specialized System One models.

\clearpage
\section*{Limitations}
SecJev evaluates finite decisions over supplied observations. Capture and experiment annotations can depend on information outside an individual input, so those tasks measure source-label prediction. Policy tasks measure application of the displayed rule. End-to-end attack prevention and autonomous response require separate deployment evaluation.

The 14 tasks have uneven difficulty: near-perfect policy scores coexist with lower vehicle-message accuracy. Task-macro averages weight these tasks equally. Recorded group boundaries address specific dependencies; residual semantic overlap and transfer beyond the tested sources remain open. Human review concentrates answer disagreements in source-label tasks, where visible telemetry provides incomplete evidence about the source annotation.

\section*{Ethical Considerations}
The corpus includes attack-bearing text and security observations for research on defensive decisions. Local aliases preserve within-scene relationships while reducing unnecessary identifiers. We retain source attribution and source-specific usage conditions rather than assigning a blanket permissive license to all underlying data. Model predictions and calibrated probabilities are fallible, particularly when evidence is incomplete; consequential actions require an application-specific policy and evaluation.

\bibliography{references}

\begin{thebibliography}{31}
\providecommand{\natexlab}[1]{#1}

\bibitem[{Almeida(2026)}]{almeida2026systemone}
Diogo Almeida. 2026.
\newblock \href
  {https://typesafe.ai/blog/introducing-system-one-models-and-jev} {Introducing
  {System One Models} \& {Jev}}.
\newblock TypeSafe AI Blog.

\bibitem[{Alsaedi et~al.(2020)Alsaedi, Moustafa, Tari, Mahmood, and
  Anwar}]{alsaedi2020ton}
Abdullah Alsaedi, Nour Moustafa, Zahir Tari, Abdun Mahmood, and Adnan Anwar.
  2020.
\newblock \href {https://doi.org/10.1109/access.2020.3022862} {{TON\_IoT
  Telemetry Dataset: A New Generation Dataset of IoT and IIoT for Data-Driven
  Intrusion Detection Systems}}.
\newblock \emph{IEEE Access}, 8:165130--165150.

\bibitem[{Ashraf et~al.(2021)Ashraf, Keshk, Moustafa, Abdel-Basset, Khurshid,
  Bakhshi, and Mostafa}]{ashraf2021iotbot}
Javed Ashraf, Marwa Keshk, Nour Moustafa, Mohamed Abdel-Basset, Hasnat
  Khurshid, Asim~D. Bakhshi, and Reham~R. Mostafa. 2021.
\newblock \href {https://doi.org/10.1016/j.scs.2021.103041} {{IoTBoT-IDS: A
  novel statistical learning-enabled botnet detection framework for protecting
  networks of smart cities}}.
\newblock \emph{Sustainable Cities and Society}, 72:103041.

\bibitem[{Bano et~al.(2022)Bano, Sonnino, Chursin, Perelman, Li, Ching, and
  Malkhi}]{bano_et_al:LIPIcs.OPODIS.2021.7}
Shehar Bano, Alberto Sonnino, Andrey Chursin, Dmitri Perelman, Zekun Li, Avery
  Ching, and Dahlia Malkhi. 2022.
\newblock \href {https://doi.org/10.4230/LIPIcs.OPODIS.2021.7} {{Twins: BFT
  Systems Made Robust}}.
\newblock In \emph{25th International Conference on Principles of Distributed
  Systems (OPODIS 2021)}, volume 217 of \emph{Leibniz International Proceedings
  in Informatics (LIPIcs)}, pages 7:1--7:29, Dagstuhl, Germany. Schloss
  Dagstuhl -- Leibniz-Zentrum f{\"u}r Informatik.

\bibitem[{Booij et~al.(2022)Booij, Chiscop, Meeuwissen, Moustafa, and {den
  Hartog}}]{booij2022ton}
Tim~M. Booij, Irina Chiscop, Erik Meeuwissen, Nour Moustafa, and Frank T.~H.
  {den Hartog}. 2022.
\newblock \href {https://doi.org/10.1109/JIOT.2021.3085194} {{ToN\_IoT}: The
  role of heterogeneity and the need for standardization of features and attack
  types in {IoT} network intrusion data sets}.
\newblock \emph{IEEE Internet of Things Journal}, 9(1):485--496.

\bibitem[{Debenedetti et~al.(2024)Debenedetti, Zhang, Balunovi{\'c},
  Beurer-Kellner, Fischer, and Tram{\`e}r}]{debenedetti2024agentdojo}
Edoardo Debenedetti, Jie Zhang, Mislav Balunovi{\'c}, Luca Beurer-Kellner, Marc
  Fischer, and Florian Tram{\`e}r. 2024.
\newblock \href {https://doi.org/10.52202/079017-2636} {{AgentDojo}: A dynamic
  environment to evaluate prompt injection attacks and defenses for {LLM}
  agents}.
\newblock In \emph{Advances in Neural Information Processing Systems},
  volume~37, pages 82895--82920. Curran Associates, Inc.

\bibitem[{Fung et~al.(2020)Fung, Yoon, and Beschastnikh}]{Fung2020}
Clement Fung, Chris J.~M. Yoon, and Ivan Beschastnikh. 2020.
\newblock \href {https://www.usenix.org/conference/raid2020/presentation/fung}
  {The limitations of federated learning in sybil settings}.
\newblock In \emph{23rd International Symposium on Research in Attacks,
  Intrusions and Defenses (RAID 2020)}, pages 301--316, San Sebastian. USENIX
  Association.

\bibitem[{Gonz{\'a}lez et~al.(2025)Gonz{\'a}lez, Guerraoui, Pinot, Rizk,
  Stephan, and Ta{\"i}ani}]{gonzalez2025byzfl}
Marc Gonz{\'a}lez, Rachid Guerraoui, Rafael Pinot, Geovani Rizk, John Stephan,
  and Fran{\c c}ois Ta{\"i}ani. 2025.
\newblock \href {https://arxiv.org/abs/2505.24802} {{ByzFL}: Research framework
  for robust federated learning}.
\newblock \emph{Preprint}, arXiv:2505.24802.

\bibitem[{Guo et~al.(2017)Guo, Pleiss, Sun, and Weinberger}]{pmlr-v70-guo17a}
Chuan Guo, Geoff Pleiss, Yu~Sun, and Kilian~Q. Weinberger. 2017.
\newblock \href {https://proceedings.mlr.press/v70/guo17a.html} {On calibration
  of modern neural networks}.
\newblock In \emph{Proceedings of the 34th International Conference on Machine
  Learning}, volume~70 of \emph{Proceedings of Machine Learning Research},
  pages 1321--1330. PMLR.

\bibitem[{Gururangan et~al.(2020)Gururangan, Marasovi{\'c}, Swayamdipta, Lo,
  Beltagy, Downey, and Smith}]{gururangan-etal-2020-dont}
Suchin Gururangan, Ana Marasovi{\'c}, Swabha Swayamdipta, Kyle Lo, Iz~Beltagy,
  Doug Downey, and Noah~A. Smith. 2020.
\newblock \href {https://doi.org/10.18653/v1/2020.acl-main.740} {Don{'}t stop
  pretraining: Adapt language models to domains and tasks}.
\newblock In \emph{Proceedings of the 58th Annual Meeting of the Association
  for Computational Linguistics}, pages 8342--8360, Online. Association for
  Computational Linguistics.

\bibitem[{Hermann et~al.(2026)Hermann, Remmers, Eisermann, Erb, and
  Kargl}]{hermann2026veremi}
Artur Hermann, Jan-Niklas Remmers, Dennis Eisermann, Benjamin Erb, and Frank
  Kargl. 2026.
\newblock \href {https://doi.org/10.5281/zenodo.19665762} {{VeReMi NextGen}: A
  dataset for evaluating misbehavior detection systems in {VANETs}}.
\newblock Zenodo, version 1.

\bibitem[{Hu et~al.(2022)Hu, Shen, Wallis, Allen-Zhu, Li, Wang, Wang, and
  Chen}]{hu2022lora}
Edward~J. Hu, Yelong Shen, Phillip Wallis, Zeyuan Allen-Zhu, Yuanzhi Li, Shean
  Wang, Lu~Wang, and Weizhu Chen. 2022.
\newblock \href {https://openreview.net/forum?id=nZeVKeeFYf9} {{LoRA}: Low-rank
  adaptation of large language models}.
\newblock In \emph{International Conference on Learning Representations}.

\bibitem[{Jing et~al.(2024)Jing, Tang, Shi, Zheng, Nie, Wu, Yang, and
  Luo}]{jing2024secbench}
Pengfei Jing, Mengyun Tang, Xiaorong Shi, Xing Zheng, Sen Nie, Shi Wu, Yong
  Yang, and Xiapu Luo. 2024.
\newblock \href {https://arxiv.org/abs/2412.20787} {{SecBench}: A comprehensive
  multi-dimensional benchmarking dataset for {LLMs} in cybersecurity}.
\newblock \emph{Preprint}, arXiv:2412.20787.

\bibitem[{Kent(2015)}]{kent-2015-cyberdata1}
Alexander~D. Kent. 2015.
\newblock \href {https://doi.org/10.17021/1179829} {Comprehensive, multi-source
  cyber-security events data set}.
\newblock Los Alamos National Laboratory.

\bibitem[{Kent(2016)}]{akent-2015-enterprise-data}
Alexander~D. Kent. 2016.
\newblock \href {https://doi.org/10.1142/9781786340757_0002} {Cyber security
  data sources for dynamic network research}.
\newblock In Niall Adams and Nick Heard, editors, \emph{Dynamic Networks and
  Cyber-Security}, pages 37--65. World Scientific.

\bibitem[{Liu et~al.(2024)Liu, Shi, and Buford}]{liu2024cyberbench}
Zefang Liu, Jialei Shi, and John~F. Buford. 2024.
\newblock \href
  {https://zefang-liu.github.io/files/liu2024cyberbench_paper.pdf}
  {{CyberBench}: A multi-task benchmark for evaluating large language models in
  cybersecurity}.
\newblock In \emph{AAAI 2024 Workshop on Artificial Intelligence for Cyber
  Security}.

\bibitem[{Moustafa(2019)}]{moustafa2019ton}
Nour Moustafa. 2019.
\newblock \href
  {https://conference.eresearch.edu.au/wp-content/uploads/2019/08/2019_eResearch_59_New-Generations-of-Internet-of-Things-Datasets-for-Cybersecurity.pdf}
  {New generations of internet of things datasets for cybersecurity
  applications based machine learning: {TON\_IoT} datasets}.
\newblock Presentation at eResearch Australasia.

\bibitem[{Moustafa(2021{\natexlab{a}})}]{moustafa2021network}
Nour Moustafa. 2021{\natexlab{a}}.
\newblock \href {https://doi.org/10.1016/j.scs.2021.102994} {{A new distributed
  architecture for evaluating AI-based security systems at the edge: Network
  TON\_IoT datasets}}.
\newblock \emph{Sustainable Cities and Society}, 72:102994.

\bibitem[{Moustafa(2021{\natexlab{b}})}]{moustafa2019fog}
Nour Moustafa. 2021{\natexlab{b}}.
\newblock \href {https://doi.org/10.1201/9781003028635-4} {A systemic
  {IoT}--fog--cloud architecture for big-data analytics and cyber security
  systems: A review of fog computing}.
\newblock In Mohiuddin Ahmed and Paul Haskell-Dowland, editors, \emph{Secure
  Edge Computing: Applications, Techniques and Challenges}, pages 41--50. CRC
  Press.

\bibitem[{Moustafa et~al.(2020{\natexlab{a}})Moustafa, Ahmed, and
  Ahmed}]{moustafa2020linux}
Nour Moustafa, Mohiuddin Ahmed, and Sherif Ahmed. 2020{\natexlab{a}}.
\newblock \href {https://doi.org/10.1109/TrustCom50675.2020.00100} {Data
  analytics-enabled intrusion detection: Evaluations of {ToN\_IoT} {Linux}
  datasets}.
\newblock In \emph{2020 IEEE 19th International Conference on Trust, Security
  and Privacy in Computing and Communications (TrustCom)}, pages 727--735.
  IEEE.

\bibitem[{Moustafa et~al.(2020{\natexlab{b}})Moustafa, Keshk, Debie, and
  Janicke}]{moustafa2020windows}
Nour Moustafa, Marwa Keshk, Essam Debie, and Helge Janicke. 2020{\natexlab{b}}.
\newblock \href {https://doi.org/10.1109/TrustCom50675.2020.00114} {Federated
  {TON\_IoT} {Windows} datasets for evaluating {AI}-based security
  applications}.
\newblock In \emph{2020 IEEE 19th International Conference on Trust, Security
  and Privacy in Computing and Communications (TrustCom)}, pages 848--855.
  IEEE.

\bibitem[{Neto et~al.(2023)Neto, Dadkhah, Ferreira, Zohourian, Lu, and
  Ghorbani}]{neto2023ciciot}
Euclides Carlos~Pinto Neto, Sajjad Dadkhah, Raphael Ferreira, Alireza
  Zohourian, Rongxing Lu, and Ali~A. Ghorbani. 2023.
\newblock \href {https://doi.org/10.3390/s23135941} {{CICIoT2023: A Real-Time
  Dataset and Benchmark for Large-Scale Attacks in IoT Environment}}.
\newblock \emph{Sensors}, 23(13):5941.

\bibitem[{Palmer(2026)}]{palmer2026kev}
Jared Palmer. 2026.
\newblock \href
  {https://github.com/jaredpalmer/kev/tree/90990a5fac2995b9faa3190f7d437e84f2067768}
  {{Kev}}.
\newblock Software.

\bibitem[{{Qwen Team}(2026)}]{qwen3.5}
{Qwen Team}. 2026.
\newblock \href {https://qwen.ai/blog?id=qwen3.5} {{Qwen3.5}: Towards native
  multimodal agents}.

\bibitem[{Shi et~al.(2025)Shi, Sun, Wang, Fu, Xu, and Li}]{shi-etal-2025-llama}
Kaize Shi, Xueyao Sun, Dingxian Wang, Yinlin Fu, Guandong Xu, and Qing Li.
  2025.
\newblock \href {https://aclanthology.org/2025.coling-main.58/} {{LL}a{MA}-{E}:
  Empowering {E}-commerce authoring with object-interleaved instruction
  following}.
\newblock In \emph{Proceedings of the 31st International Conference on
  Computational Linguistics}, pages 870--885, Abu Dhabi, UAE. Association for
  Computational Linguistics.

\bibitem[{Song et~al.(2025)Song, Zhang, and Gao}]{song-etal-2025-confront}
Shuang Song, Yifei Zhang, and Neng Gao. 2025.
\newblock \href {https://aclanthology.org/2025.coling-main.574/} {Confront
  insider threat: Precise anomaly detection in behavior logs based on {LLM}
  fine-tuning}.
\newblock In \emph{Proceedings of the 31st International Conference on
  Computational Linguistics}, pages 8589--8601, Abu Dhabi, UAE. Association for
  Computational Linguistics.

\bibitem[{Sood and Savelka(2026)}]{sood2026punching}
Bhavesh Sood and Jaromir Savelka. 2026.
\newblock \href {https://arxiv.org/abs/2607.03801} {Punching above their
  weight: Classification-head fine-tuning of tiny language models ({TLMs}) for
  verifiable multiple-choice tasks}.
\newblock \emph{Preprint}, arXiv:2607.03801.

\bibitem[{Wadden et~al.(2025)Wadden, Shi, Morrison, Li, Naik, Singh, Barzilay,
  Lo, Hope, Soldaini, Shen, Downey, Hajishirzi, and
  Cohan}]{wadden-etal-2025-sciriff}
David Wadden, Kejian Shi, Jacob Morrison, Alan Li, Aakanksha Naik, Shruti
  Singh, Nitzan Barzilay, Kyle Lo, Tom Hope, Luca Soldaini, Shannon~Zejiang
  Shen, Doug Downey, Hannaneh Hajishirzi, and Arman Cohan. 2025.
\newblock \href {https://doi.org/10.18653/v1/2025.emnlp-main.310} {{S}ci{RIFF}:
  A resource to enhance language model instruction-following over scientific
  literature}.
\newblock In \emph{Proceedings of the 2025 Conference on Empirical Methods in
  Natural Language Processing}, pages 6072--6109, Suzhou, China. Association
  for Computational Linguistics.

\bibitem[{Wu and Lim(2026)}]{wu2026decision}
Tiantong Wu and Wei Yang~Bryan Lim. 2026.
\newblock \href {https://arxiv.org/abs/2609.28613} {Decision hijacking: Prompt
  injection attacks on {Jev}'s typed probabilistic decisions}.
\newblock \emph{Preprint}, arXiv:2609.28613.

\bibitem[{Zeng et~al.(2024)Zeng, Liu, Lu, Wang, Liu, Dong, and
  Tang}]{zeng-etal-2024-agenttuning}
Aohan Zeng, Mingdao Liu, Rui Lu, Bowen Wang, Xiao Liu, Yuxiao Dong, and Jie
  Tang. 2024.
\newblock \href {https://doi.org/10.18653/v1/2024.findings-acl.181}
  {{A}gent{T}uning: Enabling generalized agent abilities for {LLM}s}.
\newblock In \emph{Findings of the Association for Computational Linguistics:
  ACL 2024}, pages 3053--3077, Bangkok, Thailand. Association for Computational
  Linguistics.

\bibitem[{Zhan et~al.(2024)Zhan, Liang, Ying, and
  Kang}]{zhan-etal-2024-injecagent}
Qiusi Zhan, Zhixiang Liang, Zifan Ying, and Daniel Kang. 2024.
\newblock \href {https://doi.org/10.18653/v1/2024.findings-acl.624}
  {{I}njec{A}gent: Benchmarking indirect prompt injections in tool-integrated
  large language model agents}.
\newblock In \emph{Findings of the Association for Computational Linguistics:
  ACL 2024}, pages 10471--10506, Bangkok, Thailand. Association for
  Computational Linguistics.

\end{thebibliography}
\clearpage
\appendix
\setcounter{dbltopnumber}{4}
\setcounter{totalnumber}{6}
\renewcommand{\dbltopfraction}{0.9}
\renewcommand{\dblfloatpagefraction}{0.8}
\section{Implementation and Model Selection}
\label{app:implementation}
Training runs span three security epochs and 118,818 questions per epoch. We train one run per condition. AdamW uses weight decay 0.01 and gradient clipping at 1.0. The first epoch and the two-epoch continuation each use a 5\% warm-up followed by cosine decay to one tenth of their respective peak learning rates; optimizer states continue across epochs. The effective batches correspond to 929 updates in epoch 1 and 619 updates in each continuation epoch, totaling 2,167 updates and 356,454 question presentations. The reported 9B epoch-2 checkpoint has 1,548 updates and 237,636 question presentations. No security evidence is truncated to fit the 8,192-token limit. The 9B run keeps the frozen backbone in BF16 and its adapters, decision head, and optimizer states in FP32; the smaller runs use FP32 master weights with BF16 autocast during training. Evaluation precision follows Section~\ref{sec:setup}.

The 2B initialization is trained on Kev's general-decision training split (upstream \texttt{decision-v7}), with the date-reasoning and missing-evidence supplement. Official Kev initializations are used at 0.8B, 4B, and 9B. The direct-Qwen control starts without this general decision stage.

\subsection{Model Selection}
\label{app:checkpoint-choice}
The 0.8B, 2B, and 4B models use epoch 3, selected by development task-macro balanced accuracy over semantic classes. The reported 9B model uses epoch 2.

\section{Corpus and Evaluation Audit}
\label{app:audit}
Table~\ref{tab:data-audit} summarizes construction checks. Two reviewers with AI and cybersecurity backgrounds independently assess 280 label-stratified questions, 20 per task and 70 per split, spanning 278 scenes. They read the complete state, question, and candidates without reference labels or model predictions, recording an answer, rationale, clarity, option exclusivity, and evidence sufficiency. Answers may include ``insufficient information'' or ``ambiguous'' (Table~\ref{tab:human-review}).

\paragraph{Answer agreement.}
The reviewers agree on 261 of 280 answers (93.21\%). All 180 tool-text and explicit-policy answers agree between reviewers and match the corpus labels. The 100 source-label questions yield 81 agreements: 20/20 for CICIoT, 17/20 for ByzFL, 11/20 for ToN-IoT, 18/20 for VeReMi manipulation, and 15/20 for VeReMi attack type. Reviewers A and C match the corpus labels on 250/280 and 261/280 questions, respectively. Fourteen shared answers differ from the source labels. C records insufficient information for two VeReMi questions; these remain in the 280-question denominator.

\paragraph{Evidence quality.}
Both reviewers judge all 280 questions clear. A judges all option sets mutually exclusive; C marks two VeReMi attack-type sets uncertain. Evidence-sufficiency judgments agree on 244 questions (87.14\%). A rates 232 as sufficient and 48 as partial; C rates 207 as sufficient, 58 as partial, and 15 as insufficient. Joint sufficiency covers 179/180 tool-text and explicit-policy questions and 28/100 source-label questions. The task profile separates applying a stated policy or interpreting a tool response from recovering an experiment or capture annotation. Review rationales identify short vehicle histories, unknown absolute positions, and network or update patterns compatible with multiple source conditions as recurring causes of uncertainty.

Source rechecks cover 20 original InjecAgent records, 30 network CSV rows, and 20 retained ByzFL construction records; retained fields and label mappings match. VeReMi checks cover temporal structure, target selection, and binary/type consistency, rather than independently establishing hidden simulator labels. One InjecAgent source example exposes an operation-category ambiguity: updating profile fields is labeled \texttt{other\_operation}, while \texttt{information\_transfer} includes entering information on a website.
\begin{table*}[tp]
\centering
\small
\begin{tabular*}{\textwidth}{@{}@{\extracolsep{\fill}}p{0.29\textwidth}p{0.26\textwidth}p{0.36\textwidth}@{}}
\toprule
Check & Scope & Result \\
\midrule
Schema and label validity & 170,185 questions & No violations found \\
Scene duplication and split isolation & 115,324 scenes and source groups & No violations found under the recorded grouping rules \\
Independent policy recomputation & 37,490 LANL/Twins questions & 0 inconsistent answers \\
Semantic-class distribution & Every task and split & Maximum class share 80\%. Maximum cross-split gap 8.58 pp \\
Source-record recheck & 50 original records; 20 retained ByzFL records & No mismatches in the checked fields or label mappings \\
Independent human review & 280 questions; 20 per task & 261 answer agreements (93.21\%); 180/180 tool-text and policy answers match labels \\
Source-label observability & 100 source-classification questions & 81 answer agreements; 28 rated sufficient by both reviewers \\
\bottomrule
\end{tabular*}
\caption{Corpus construction checks, source-record rechecks, and independent human review.}
\label{tab:data-audit}
\end{table*}

\begin{table*}[tp]
\centering\small
\begin{tabular*}{\textwidth}{@{}@{\extracolsep{\fill}}lrrrrrr@{}}
\toprule
 & & & \multicolumn{2}{c}{Matches corpus label} & \multicolumn{2}{c}{Evidence sufficient} \\
\cmidrule(lr){4-5}\cmidrule(lr){6-7}
Task & Questions & A--C agree & A & C & A & C \\
\midrule
AgentDojo: boundary action & 20 & 20 & 20 & 20 & 20 & 19 \\
AgentDojo: injection & 20 & 20 & 20 & 20 & 20 & 20 \\
CICIoT: attack & 20 & 20 & 14 & 14 & 6 & 3 \\
ByzFL: update attack & 20 & 17 & 17 & 18 & 4 & 0 \\
InjecAgent: operation & 20 & 20 & 20 & 20 & 20 & 20 \\
LANL: logon action & 20 & 20 & 20 & 20 & 20 & 20 \\
LANL: failure threshold & 20 & 20 & 20 & 20 & 20 & 20 \\
ToN-IoT: attack & 20 & 11 & 11 & 20 & 3 & 0 \\
Twins: evidence level & 20 & 20 & 20 & 20 & 20 & 20 \\
Twins: policy action & 20 & 20 & 20 & 20 & 20 & 20 \\
Twins: quorum & 20 & 20 & 20 & 20 & 20 & 20 \\
Twins: visible conflict & 20 & 20 & 20 & 20 & 20 & 20 \\
VeReMi: manipulation & 20 & 18 & 14 & 15 & 19 & 11 \\
VeReMi: attack type & 20 & 15 & 14 & 14 & 20 & 14 \\
\midrule
Tool text and explicit policy & 180 & 180 & 180 & 180 & 180 & 179 \\
Source-label prediction & 100 & 81 & 70 & 81 & 52 & 28 \\
All tasks & 280 & 261 & 250 & 261 & 232 & 207 \\
\bottomrule
\end{tabular*}
\caption{Independent human review of 280 questions. Entries are counts. A--C agreement compares the reviewers' answers; label matches compare each answer with the corpus target. Evidence sufficiency is a separate judgment.}
\label{tab:human-review}
\end{table*}

\section{Semantic Consistency by Task}
\label{app:robust-details}
Table~\ref{tab:robust-tasks} separates original, paraphrased, and reordered accuracy by task. The original column for a choice task is also its permutation reference, since all 100 questions are shared. Each question has a single paraphrase and each eligible choice has a single permutation. Across 0.8B, 2B, and 4B, paraphrasing changes 38, 28, and 15 correct predictions to incorrect ones; permutation does so for 4, 7, and 4 predictions, respectively. For 9B epoch 2, paraphrasing introduces 25 errors and corrects 24; permutation introduces two and corrects four. Reverse transitions partially offset these losses in aggregate accuracy.

Batch-one checks reproduce the smaller-model predictions, including all 163 checked discordant pairs across 0.8B, 2B, and 4B. The reported 9B epoch-2 checkpoint is evaluated on all 3,300 requests in fixed batches ordered by full-suite length and identifier.

\begin{table*}[tp]
\centering\small
\begin{tabular*}{\textwidth}{@{}@{\extracolsep{\fill}}lrrrr@{}}
\toprule
Task & 0.8B & 2B & 4B & 9B \\
\midrule
AgentDojo: boundary action & 78 / 81 / 79 & 91 / 90 / 88 & 99 / 99 / 99 & 97 / 97 / 98 \\
AgentDojo: injection & 85 / 86 / -- & 88 / 91 / -- & 95 / 95 / -- & 98 / 98 / -- \\
CICIoT: attack & 93 / 94 / -- & 94 / 93 / -- & 93 / 94 / -- & 88 / 95 / -- \\
ByzFL: update attack & 91 / 85 / -- & 95 / 94 / -- & 88 / 84 / -- & 83 / 85 / -- \\
InjecAgent: operation & 100 / 98 / 100 & 98 / 94 / 98 & 100 / 100 / 100 & 100 / 99 / 100 \\
LANL: logon action & 100 / 100 / 100 & 100 / 100 / 100 & 100 / 100 / 100 & 99 / 100 / 99 \\
LANL: failure threshold & 100 / 100 / -- & 100 / 100 / -- & 100 / 100 / -- & 100 / 100 / -- \\
ToN-IoT: attack & 94 / 95 / -- & 96 / 95 / -- & 95 / 95 / -- & 95 / 89 / -- \\
Twins: evidence level & 99 / 98 / -- & 100 / 100 / -- & 100 / 100 / -- & 100 / 100 / -- \\
Twins: policy action & 98 / 98 / 97 & 100 / 100 / 100 & 100 / 100 / 100 & 100 / 100 / 99 \\
Twins: quorum & 100 / 99 / -- & 100 / 99 / -- & 100 / 98 / -- & 100 / 97 / -- \\
Twins: visible conflict & 97 / 99 / -- & 98 / 98 / -- & 100 / 98 / -- & 100 / 100 / -- \\
VeReMi: manipulation & 67 / 64 / -- & 76 / 68 / -- & 67 / 68 / -- & 69 / 68 / -- \\
VeReMi: attack type & 39 / 32 / 37 & 35 / 36 / 34 & 46 / 45 / 46 & 41 / 41 / 43 \\
\bottomrule
\end{tabular*}
\caption{Accuracy by task (\%): original / paraphrase / permutation, with 100 questions per applicable condition. A dash means permutation is not evaluated for that answer type.}
\label{tab:robust-tasks}
\end{table*}

\section{Direct-Qwen Initialization}
\label{app:qwen-init}
The 0.8B initialization comparison fixes the security examples and their order, learning-rate schedule, effective batches, and scene-weighted loss. Both runs select epoch 3 using development task-macro balanced accuracy. Evaluation uses FP32 backbones, adapters, and heads. The first epoch uses two devices for Kev initialization and three for direct-Qwen initialization at the same effective batch; subsequent epochs use three devices. Table~\ref{tab:qwen-tasks} gives task-level results for the selected checkpoints.

\begin{table*}[tp]
\centering\small
\begin{tabular*}{\textwidth}{@{}@{\extracolsep{\fill}}lrrr@{}}
\toprule
Task & Questions & Kev initialization & Direct-Qwen initialization \\
\midrule
AgentDojo: boundary action & 426 & 80.28 & 82.86 \\
AgentDojo: injection & 426 & 79.81 & 71.60 \\
ByzFL: update attack & 800 & 92.62 & 95.38 \\
CICIoT: attack & 6000 & 91.80 & 91.47 \\
InjecAgent: operation & 124 & 100.00 & 96.77 \\
LANL: logon action & 540 & 98.89 & 97.78 \\
LANL: failure threshold & 540 & 99.81 & 99.44 \\
ToN-IoT: attack & 3500 & 98.03 & 98.11 \\
Twins: evidence level & 659 & 98.79 & 98.63 \\
Twins: policy action & 659 & 98.48 & 96.81 \\
Twins: quorum & 659 & 99.70 & 97.57 \\
Twins: visible conflict & 659 & 99.24 & 98.94 \\
VeReMi: manipulation & 3321 & 63.11 & 54.05 \\
VeReMi: attack type & 3321 & 64.14 & 54.95 \\
\bottomrule
\end{tabular*}
\caption{Per-task test accuracy for the development-selected 0.8B checkpoints (\%).}
\label{tab:qwen-tasks}
\end{table*}

\section{Paired Uncertainty and General Confidence}
\label{app:statistics}
The training-majority baseline selects a semantic class separately for each task from 118,818 training questions; deterministic lexical order resolves ties. On the 21,634 test questions, it scores 64.25\% micro accuracy, 62.89\% task-macro accuracy, and 43.57\% task-macro balanced accuracy.

For paired intervals, we resample whole groups within each source, keeping all questions from a selected group together and applying the same multiplicity across tasks. We use 2,000 draws with seed 20260929 and report the 2.5th and 97.5th percentiles. Every draw includes all tasks. The security test contains 485 groups: AgentDojo 2, ByzFL 5, CICIoT2023 8, InjecAgent 2, LANL 7, ToN-IoT 6, Twins 34, and VeReMi 421. The two general suites contain 880 and 544 groups, respectively. These intervals describe variation over the observed source groups; estimates are coarser for sources with few groups.

\begin{table*}[tp]
\centering\small
\begin{tabular*}{\textwidth}{@{}@{\extracolsep{\fill}}lrrr@{}}
\toprule
Size & Security task-macro & General decision micro & General transfer micro \\
\midrule
0.8B & $+39.40$ $[+30.16, +43.48]$ & $-0.42$ $[-1.42, +0.58]$ & $-2.44$ $[-4.73, -0.15]$ \\
2B & $+40.83$ $[+28.62, +45.32]$ & $+0.50$ $[-0.50, +1.42]$ & $+0.00$ $[-1.98, +1.98]$ \\
4B & $+26.63$ $[+18.59, +30.75]$ & $-0.50$ $[-1.17, +0.17]$ & $+0.15$ $[-1.07, +1.37]$ \\
9B & $+24.26$ $[+15.38, +28.40]$ & $-0.08$ $[-0.58, +0.42]$ & $-0.46$ $[-2.29, +1.37]$ \\
\bottomrule
\end{tabular*}
\caption{SecJev minus initialization, in percentage points, with paired 95\% group-bootstrap intervals. Positive differences favor SecJev. These are intervals for accuracy differences, not for absolute scores.}
\label{tab:paired-intervals}
\end{table*}

The general decision and general transfer suites use Kev's upstream versions \texttt{decision-v7} and \texttt{transfer-v4}, respectively; the suffixes identify dataset revisions. We evaluate their clean subsets of 1,200 and 656 questions (Table~\ref{tab:general-confidence}). Raw probabilities use $T=1$. Scaled SecJev probabilities use the temperature fitted on security calibration data. Initialization rows retain their original stored temperatures; the 2B initialization uses $T=1$. Thus the raw columns compare confidence without temperature differences, whereas the scaled columns describe the stored models as used. NLL averages negative log probability of the correct candidate; multiclass Brier sums squared probability error across candidates; ECE uses 15 equal-width confidence bins.

All SecJev models have higher raw NLL than their initialization on both general suites, despite the small accuracy changes. Security temperature scaling lowers all three general-suite metrics for every SecJev size, but the calibrated 0.8B transfer ECE remains 21.10\%, compared with 4.19\% for its initialization. General decision accuracy and transferable confidence are therefore distinct aspects of retention.

\begin{table*}[tp]
\centering\small
\setlength{\tabcolsep}{3pt}
\begin{tabular*}{\textwidth}{@{}@{\extracolsep{\fill}}lllrrrrrrr@{}}
\toprule
& & & & \multicolumn{3}{c}{Raw ($T=1$)} & \multicolumn{3}{c}{Stored temperature} \\
\cmidrule(lr){5-7}\cmidrule(lr){8-10}
Suite & Size & Model & $T$ & NLL & Brier & ECE & NLL & Brier & ECE \\
\midrule
General decision & 0.8B & Init. & 2.406 & 0.652 & 0.268 & 9.94 & 0.449 & 0.240 & 2.69 \\
General decision & 0.8B & SecJev & 1.353 & 0.946 & 0.293 & 12.86 & 0.734 & 0.281 & 11.26 \\
General decision & 2B & Init. & 1.000 & 0.574 & 0.241 & 9.22 & 0.574 & 0.241 & 9.22 \\
General decision & 2B & SecJev & 1.292 & 0.749 & 0.249 & 9.88 & 0.616 & 0.241 & 8.48 \\
General decision & 4B & Init. & 2.144 & 0.468 & 0.207 & 7.32 & 0.360 & 0.189 & 2.60 \\
General decision & 4B & SecJev & 1.335 & 0.573 & 0.218 & 8.83 & 0.465 & 0.208 & 7.31 \\
General decision & 9B & Init. & 2.297 & 0.479 & 0.201 & 7.28 & 0.352 & 0.181 & 1.90 \\
General decision & 9B & SecJev & 1.181 & 0.533 & 0.206 & 7.57 & 0.472 & 0.201 & 6.81 \\
\midrule
General transfer & 0.8B & Init. & 2.406 & 0.868 & 0.460 & 15.63 & 0.692 & 0.412 & 4.19 \\
General transfer & 0.8B & SecJev & 1.353 & 1.357 & 0.543 & 24.36 & 1.079 & 0.511 & 21.10 \\
General transfer & 2B & Init. & 1.000 & 0.687 & 0.343 & 11.58 & 0.687 & 0.343 & 11.58 \\
General transfer & 2B & SecJev & 1.292 & 0.941 & 0.382 & 15.02 & 0.787 & 0.365 & 12.50 \\
General transfer & 4B & Init. & 2.144 & 0.564 & 0.255 & 10.00 & 0.426 & 0.232 & 5.00 \\
General transfer & 4B & SecJev & 1.335 & 0.841 & 0.277 & 11.57 & 0.671 & 0.267 & 9.16 \\
General transfer & 9B & Init. & 2.297 & 0.566 & 0.238 & 8.38 & 0.417 & 0.224 & 4.02 \\
General transfer & 9B & SecJev & 1.181 & 0.593 & 0.227 & 9.45 & 0.528 & 0.221 & 8.39 \\
\bottomrule
\end{tabular*}
\caption{General-suite confidence before and after security specialization. Lower is better; ECE is in percent. No temperature is fitted on these general test subsets. Init. is Kev at 0.8B/4B/9B and the general-decision starting point at 2B.}
\label{tab:general-confidence}
\end{table*}

\section{Generative and Classification Baselines}
\label{app:generative}
SecGen-0.8B and SecGen-4B start from Qwen3.5 and train rank-16 LoRA adapters on the same 118,818 security questions for three epochs, each totaling 356,454 question presentations and 2,167 updates. The original backbone and vocabulary head are frozen. Training retains FP32 weights and optimizer states with BF16 autocast; development, calibration, and test inference use FP32 throughout, without TF32, at batch size one. Supervision is full-vocabulary cross-entropy on one answer-key token, weighted by the inverse number of questions in its scene. Decoding selects the highest-scoring valid key and ends after that token. For these one-token keys, ranking their vocabulary log-probabilities is identical to constrained greedy selection.

The 0.8B candidate-sequence classifier adapts \citet{sood2026punching}: each state and question is concatenated with one candidate, and its final nonpadding hidden state is mapped to a scalar by a shared bias-free linear head. Cross-entropy normalizes these scores across the question's candidates. Starting from Qwen3.5-0.8B-Base, we train the fresh head and rank-16 LoRA adapters with the same security data, scene weighting, optimizer schedule, and three-epoch budget as the other Qwen controls. Training uses BF16 autocast with FP32 master parameters; evaluation is FP32 without TF32. The three epochs take 8.75 hours on nine RTX 3090s. This comparison evaluates a candidate-conditioned encoding interface, which repeats the context for each option, alongside SecJev's shared encoding of all candidates.

Each epoch is evaluated on all 14,912 development questions. Semantic-class balanced macro selects epoch 3 for SecGen at both sizes (Table~\ref{tab:generative-dev}); ties select the earlier epoch. Temperatures of 1.466 and 1.264 for 0.8B and 4B are fitted to candidate-normalized NLL on 14,821 calibration questions, before evaluating the 21,634 test questions. At both sizes, the unconstrained vocabulary argmax is a valid answer key for every test question. For classification, full-development semantic-class balanced macro selects epoch 3; the same calibration split gives a single global temperature of 1.288.
\begin{table}[!t]
\centering\small
\begin{tabular*}{\columnwidth}{@{}@{\extracolsep{\fill}}llrrr@{}}
\toprule
Model & Epoch & Micro & Macro & Bal. macro \\
\midrule
SecGen-0.8B & 3 & 89.40 & 91.90 & 89.46 \\
\midrule
Cls-0.8B & 3 & 89.09 & 90.76 & 88.39 \\
\midrule
SecGen-4B & 3 & 91.15 & 94.11 & 92.35 \\
\bottomrule
\end{tabular*}
\caption{Development accuracy (\%) for selected SecGen and candidate-sequence classification (Cls) checkpoints. Balanced macro groups recall by semantic answer label.}
\label{tab:generative-dev}
\end{table}

\begin{table*}[tp]
\centering\small
\begin{tabular*}{\textwidth}{@{}@{\extracolsep{\fill}}llrrr@{}}
\toprule
Size & Method & Test micro & Test macro & Test balanced macro \\
\midrule
0.8B & SecJev (Kev initialization) & 85.03 & 90.34 & 88.00 \\
0.8B & Direct-Qwen pointer & 81.96 & 88.17 & 83.75 \\
0.8B & SecGen & 84.75 & 90.02 & 87.79 \\
0.8B & Candidate-sequence classification & 84.39 & 89.85 & 86.84 \\
\midrule
4B & SecJev (Kev initialization) & 88.01 & 94.51 & 92.42 \\
4B & SecGen & 87.95 & 94.35 & 92.30 \\
\bottomrule
\end{tabular*}
\caption{FP32 test accuracy for development-selected models (\%). All select epoch 3. These scores use FP32; Table~\ref{tab:efficiency} uses a BF16 backbone.}
\label{tab:generative-test}
\end{table*}

Table~\ref{tab:generative-test} compares the selected models at FP32 precision. Task-level results in Table~\ref{tab:generative-tasks} locate the aggregate differences across decision interfaces: classification exceeds SecGen-0.8B on boundary action and traffic classification, while SecGen leads on injection identification, ByzFL, and VeReMi manipulation. Table~\ref{tab:efficiency} reports inference cost.

\begin{table*}[tp]
\centering\small
\begin{tabular*}{\textwidth}{@{}@{\extracolsep{\fill}}lrrrrrrr@{}}
\toprule
 & & \multicolumn{2}{c}{SecGen-0.8B} & \multicolumn{2}{c}{Cls-0.8B} & \multicolumn{2}{c}{SecGen-4B} \\
\cmidrule(lr){3-4}\cmidrule(lr){5-6}\cmidrule(lr){7-8}
Task & Questions & Acc. & Bal. & Acc. & Bal. & Acc. & Bal. \\
\midrule
AgentDojo: boundary action & 426 & 76.29 & 80.63 & 84.27 & 80.62 & 96.71 & 94.92 \\
AgentDojo: injection & 426 & 79.11 & 80.24 & 75.35 & 74.49 & 97.89 & 96.42 \\
ByzFL: update attack & 800 & 95.62 & 93.05 & 94.00 & 88.28 & 97.25 & 96.17 \\
CICIoT: attack & 6000 & 91.30 & 86.44 & 92.07 & 89.33 & 91.70 & 87.29 \\
InjecAgent: operation & 124 & 99.19 & 99.07 & 100.00 & 100.00 & 100.00 & 100.00 \\
LANL: logon action & 540 & 98.89 & 98.64 & 98.70 & 98.36 & 100.00 & 100.00 \\
LANL: failure threshold & 540 & 100.00 & 100.00 & 100.00 & 100.00 & 100.00 & 100.00 \\
ToN-IoT: attack & 3500 & 98.29 & 96.88 & 99.00 & 98.46 & 98.94 & 98.19 \\
Twins: evidence level & 659 & 98.79 & 98.67 & 97.27 & 97.00 & 99.85 & 99.83 \\
Twins: policy action & 659 & 98.63 & 98.50 & 97.42 & 97.21 & 99.70 & 99.66 \\
Twins: quorum & 659 & 99.24 & 99.12 & 99.54 & 99.55 & 99.85 & 99.88 \\
Twins: visible conflict & 659 & 98.94 & 98.38 & 98.33 & 97.66 & 99.85 & 99.75 \\
VeReMi: manipulation & 3321 & 66.06 & 66.67 & 61.88 & 62.44 & 70.58 & 71.21 \\
VeReMi: attack type & 3321 & 59.98 & 32.70 & 60.10 & 32.38 & 68.56 & 48.80 \\
\bottomrule
\end{tabular*}
\caption{Full-test results for selected SecGen and classification (Cls) checkpoints (\%). Balanced accuracy averages recall across semantic answer labels, independent of candidate order.}
\label{tab:generative-tasks}
\end{table*}

\section{Matched-Precision Timing Details}
\label{app:timing}
An answer is complete when the final structured decision is available to the caller. Timing includes tokenization and output parsing, and excludes model loading. All six models run serially on the same RTX 3090 with Python 3.11.11, PyTorch 2.6.0, CUDA 12.6, Transformers 5.17.0, PEFT 0.21.0, and NVIDIA driver 570.133.20. Both interfaces use scaled dot-product attention (SDPA). Backbone weights are BF16; original FP32 adapter and output-head values are retained. Full-test accuracy is recomputed serially at batch size one on all 21,634 questions per model (Table~\ref{tab:timing-accuracy}). The Kev-9B scores in Table~\ref{tab:security-results} use its original three-shard batched evaluation. Balanced macro groups gold answers by semantic label, not by their displayed position.

The fixed timing workload contains 200 development questions from distinct scenes, selected by hash order and task round-robin without predictions or labels. Length limits hold for every method. The 100 short requests span 11 tasks; the long stratum contains 34 ByzFL and 33 questions from each LANL task. Thus length-stratum results also reflect different task composition. Each pass uses the same requests in a shuffled order with seed $20260929+r$, for pass $r=0,\ldots,4$, following two warm-up requests per stratum. Timed predictions are not used for model selection.

Table~\ref{tab:timing-dispersion} gives the range of per-pass medians and the mean and sample standard deviation of serial batch-one throughput. In each pass, throughput is $200/\sum_{i=1}^{200} t_i$, where $t_i$ is the measured request time in seconds. The fixed workload contains equal numbers of short and long requests; throughput therefore differs from the reciprocal of either stratum median. All six models complete five serial batch-one passes.
\begin{table*}[tp]
\centering\small
\begin{tabular*}{\textwidth}{@{}@{\extracolsep{\fill}}llrrr@{}}
\toprule
Size & Method & Micro & Macro & Semantic balanced macro \\
\midrule
0.8B & SecJev & 85.14 & 90.46 & 88.04 \\
0.8B & SecGen & 84.70 & 90.01 & 87.76 \\
0.8B & Direct-Qwen pointer & 81.85 & 88.12 & 83.72 \\
4B & SecJev & 87.94 & 94.48 & 92.38 \\
4B & SecGen & 87.95 & 94.34 & 92.26 \\
9B & Kev & 50.87 & 69.79 & 70.32 \\
\bottomrule
\end{tabular*}
\caption{Full-test accuracy (\%) recomputed with the serial BF16-backbone timing implementation. Each row contains 21,634 unique questions across 14 tasks.}
\label{tab:timing-accuracy}
\end{table*}
\begin{table*}[tp]
\centering\small
\begin{tabular*}{\textwidth}{@{}@{\extracolsep{\fill}}llrrr@{}}
\toprule
Size & Method & Short median range (ms) & Long median range (ms) & Batch-one Q/s \\
\midrule
0.8B & SecJev & 79.0--79.7 & 163.5--164.4 & $7.319\pm0.016$ \\
0.8B & SecGen & 80.2--80.7 & 166.8--167.6 & $7.220\pm0.007$ \\
0.8B & Direct-Qwen pointer & 80.3--80.7 & 164.5--164.7 & $7.272\pm0.006$ \\
4B & SecJev & 107.4--110.3 & 577.6--579.2 & $2.537\pm0.001$ \\
4B & SecGen & 112.9--135.2 & 597.5--600.3 & $2.470\pm0.026$ \\
9B & Kev & 168.8--169.4 & 859.3--862.0 & $1.726\pm0.003$ \\
\bottomrule
\end{tabular*}
\caption{Variation over five timing passes. Ranges describe per-pass medians, not confidence intervals. Serial batch-one throughput is mean $\pm$ sample standard deviation over the same 200-request workload.}
\label{tab:timing-dispersion}
\end{table*}

\end{document}